\documentclass[12pt,a4paper]{article}
\usepackage{amsmath}
\usepackage{amssymb}
\usepackage{latexsym}
\usepackage{graphics}
\usepackage{psfrag,epsfig}

\begin{document}

\begin{titlepage}

\hfill hep-th/0611094\\

\vspace*{1.5cm}
\begin{center}
{\bf \Large Dark Energy From Bulk Matter}

\bigskip \bigskip \medskip

{\bf  C. Bogdanos, A. Dimitriadis and K. Tamvakis}

\bigskip

{ Physics Department, University of Ioannina\\
Ioannina GR451 10, Greece}

\bigskip \medskip
{\bf Abstract}
\end{center}
We consider the possibility of getting accelerated expansion  and $w=-1$
crossing in the context of a braneworld  cosmological setup, endowed with a bulk
energy-momentum tensor. For a given ansatz of the bulk content,
we demonstrate that the bulk pressures dominate the dynamics at late times and
can lead to accelerated expansion. We also analyze the constraints under
which we can get a realistic profile for the effective equation of state and 
conclude that matter in the bulk has the effect of dark energy on the brane. 
Furthermore, we show that it is possible to simulate the behavior of a
Chaplygin gas using non-exotic bulk matter.

\end{titlepage}

Recent experimental results suggesting an accelerating expansion of the
universe\cite{R}, \cite{WMAP}, have become the subject of active research.
During the last few years, a number of theories were proposed
to answer the question of the origin of such an acceleration. The most common
explanation is to assume the existence of some, for the moment unknown,
{\textit{``dark energy"}} component, which drives the expansion. 
The physics of this dark energy is still speculative, as no direct experiment
can provide us with actual information about its nature. We can only
indirectly probe its properties by studying the dynamics of cosmic expansion.
There are several theories which attempt to describe dark energy. Among them
are the phantom~\cite{PHANT} and 
quintessence~\cite{QUINT} theories, quiessence (cosmological constant) and
other forms of exotic matter~\cite{WEI}, like the Chaplygin gas~\cite{KMP}~(see
also \cite{SS} for a review and a more complete list of references). In these
theories, dark energy is attributed to the presence of 
new fields with non-trivially modified equations of state.

One can follow a different route and try instead modified theories of gravity. 
These are theories which extend or modify GR in such a way, so that dark energy is not 
some new -maybe exotic- matter constituent, but merely a manifestation of the modified dynamics of geometry. 
In this category we find scalar-tensor theories of gravity, the so-called modified gravity models 
and theories with extra dimensions~\cite{AHDD}, \cite{DGP}. Braneworld models,
where our four-dimensional world is realized as a brane embedded in a higher
dimensional space (bulk), have received considerable attention mainly due to the
hope of tackling hierarchy problems and explaining
the large disparity between the Planck and the electroweak scale~\cite{RS}. 
Brane models like these are also directly connected to String Theory. 
An interesting consequence of brane models is that they allow for the
presence of five-dimensional matter which propagates in the bulk space and may
interact with the matter content of the braneworld~\cite{KKTTZ}. Such an
interaction can alter the profile of the cosmic expansion and lead to a behavior
that would resemble that of dark energy.

It has already been shown~\cite{BT} that a configuration of bulk matter resembling 
a perfect fluid moving along the fifth dimension is capable of producing accelerated
expansion and give an effective equation of state which can cross the $w=-1$
line. However, for the later to occur, the existence of a large and negative
dark radiation term was needed. The presence of such a term may cause problems
in early times, as it could lead to a negative 
effective energy density (although the validity of the modified
Friedmann equation for large redshifts is in question). Our goal here is to
replicate these effects by ignoring the dark 
radiation term and considering only bulk matter as the late time dominant term.
We are also interested in determining not only under which conditions we can
have $w_{eff}<-1$, but whether
we can obtain a temporal profile of the crossing which is at least
qualitatively in accordance 
with observations for $0 \le z \le 1$. As we will see, these requirements can
be met, provided we 
have a specific ansatz for bulk matter and the parameters are properly set. We
will also demonstrate
the equivalence of this bulk matter to a Chaplygin gas.

We will start by considering a braneworld cosmological model, with a 5D bulk space and 
a positive tension brane embedded in it. The action of the model is
\begin{equation}
{\cal{S}} \,= \,\int {d^5 x\sqrt { - G} \left( {\,2M^3 R - \Lambda  +{\cal{L}}_B ^{(m)} } \right)\, + \,\int {d^4 x} \sqrt { -g} \left( { - \sigma  +
{\cal{L}}_{b} ^{(m)} } \right)} .
\end{equation}
The 5D metric is $G_{MN}$ with signature $(-,+,+,+,+)$, 
while $g_{\mu \nu}$ is the induced metric on the brane and  M denotes the five-dimensional Planck mass.
We shall assume that the metric can be written as
\begin{equation}
ds^2\,=\,-n^2(y,t)dt^2\,+\,a^2(y,t)\gamma_{ij}dx^idx^j\,+\,b^2(y,t)dy^2, \label{metric}
\end{equation}
 where $y$ denotes the fifth dimension (the brane is situated at $y=0$) and $\gamma_{ij}$ is the metric of a maximally 
 symmetric three-space. 
  
Varying the action with respect to the metric,
we obtain Einstein's equations
\begin{equation}
{\cal{G}}_{MN}\equiv\,R_{MN}  - \frac{1}
{2}G_{MN} R \,=\,\frac{1}{4M^3}\,T_{MN}\, {\label{Einstein1}}.
\end{equation}
The energy-momentum tensor $T_{MN}$ resulting from the above Action is of the form
\begin{equation}
T_{MN}\,=\,T^{( B )} _{MN} \, +\, T^{( b )} _{MN} \, - G_{MN} \Lambda  \,-g_{\mu\nu}\, \sigma \,\delta ( y
)\,\delta_M^{\mu}\delta_N^{\nu}\,{\label{e-m1}} ,
\end{equation}
where $T_{MN}^{(B)}$ results from ${\cal{L}}_B^{(m)}$ and $T_{MN}^{(b)}$ results from ${\cal{L}}_b^{(m)}$. In particular,
 we will assume an energy-momentum tensor for the bulk content of the form
\begin{equation}
{T^{(B)}}^{M}_{\,\,N}\,=\,\left(\begin{array}{ccc}
-\rho_B\,&\,0\,&\,P_5\\ 
\,0\,&\,P_B\delta^i_{\,\,j}\,&\,0\\
-\frac{n^2}{b^2}P_5\,&\,0\,&\,\overline{P}_B
\end{array}\right)\,\,,\,\,\,
T_{MN}^{(B)}\,=\,\left(\begin{array}{ccc}
\rho_Bn^2\,&\,0\,&\,-n^2P_5\\ 
\,0\,&\,P_Ba^2\gamma_{ij}\,&\,0\\
-n^2P_5\,&\,0\,&\,\overline{P}_Bb^2
\end{array}\right)\,{\label{bulk-e-m}} .
\end{equation}
The quantities which are of interest here are the pressures $\bar P_{B}$ and $P_5$, as these two enter the cosmological equations of motion. $P_5$ is the term responsible for energy exchange between the brane and the bulk. The corresponding energy-momentum tensor for the brane matter is
\begin{equation}
{T^{(b)}}^{M}_{\,\,N}\,=\,\frac{\delta(y)}{b}diag(-\rho,p,p,p,0)\,.{\label{brane-e-m}}
\end{equation}

Substituting the above ansatze for the metric (\ref{metric}) and for the energy-momentum tensor (\ref{bulk-e-m}), (\ref{brane-e-m}) into the equations of
motion (\ref{Einstein1}), we can obtain the set of cosmological equations for the components of the
metric~\cite{BDL}\cite{KKTTZ}\cite{BT}. These equations simplify if we choose
{\textit{Gauss normal coordinates}},
 such that $b(t,y)=1$. We can also make use
of the freedom to take $n(0,t)=n_0(t)=1$. After these simplifications, we get for the evolution
equations\footnote{$k=-1,\,0,\,1$ is the maximally symmetric internal space curvature parameter.} on the brane
($a_0(t)\equiv a(0,t)$)
\begin{equation}
\dot{\rho}+3\left(\,\rho+p\,\right)\frac{\dot{a}_0}{a_0}=-2P_5\,,{\label{continuity}}
\end{equation}
\begin{equation}
\frac{\ddot{a}_0}{a_0}\,+\,\left(\frac{\dot{a}_0}{a_0}\right)^2\,+\,\frac{k}{a_0^2}\,=\,\frac{1}{(24M^3)^2}(\sigma+\rho)\left(2\sigma-\rho-3p\right)\,+\,
\frac{1}{12M^3}\left(\,\Lambda\,-\overline{P}_B\,\right)\,.{\label{friedmann1}}
\end{equation}
The first expresses energy conservation on the brane. We see that positive $P_{5}$ means energy outflow from the brane, while for $P_{5}$ negative we get the opposite effect, with energy flowing from the bulk into the brane. The second equation is the Friedmann equation for the brane world. The $``55"$ component of the bulk 
pressure appears on the right hand side of this equation (\ref{friedmann1}) affecting
the cosmological evolution.

We assume that matter on the brane has an equation of state $\rho=wp$. 
Introducing the parameters  $\beta \equiv (24M^3)^{-2}$ and $\gamma \equiv
\sigma\beta$ and omitting the ``o" subscript from the scale factor, we can
rewrite the cosmological evolution equation (\ref{friedmann1}) as
\begin{equation}
\frac{\ddot{a}}{a}\,+\,\left(\frac{\dot{a}}{a}\right)^2\,+\,\frac{k}{a^2}\,=
\,\gamma\rho(1-3w)-\beta\rho^2(1+3w)-\frac{\overline{P}_B}{12M^3}+\frac{\lambda}{12M^3}\,.{\label{friedmann2}}
\end{equation}
We also assume the Randall-Sundrum fine-tuning to hold, so that the effective cosmological constant on the brane
\begin{equation}
\lambda\equiv \,\Lambda\,+\,\frac{\sigma^2}{24M^3}\,,
\end{equation}
is equal to zero, i.e. $\lambda=0$. The cosmological evolution equation (\ref{friedmann2}) can
 then be expressed as a set of two equations using a {\textit{dark energy field}} $\chi$ 
\begin{equation}
\left( {\frac{{\dot a}}
{a}} \right)^2 \, =\,\beta \rho ^2  + 2\gamma \rho  - \frac{k}
{{a^2 }} + \chi  - \frac{\overline{P}_B }
{{12M^{3}}}\,,\label{feq}
\end{equation}
\begin{equation}
\dot \chi  + 4\frac{{\dot a}}
{a}\left( \chi  - \frac{\overline{P}_B }
{24M^{3}} \right)\,=\,4\beta \left( {\rho  + \frac{\gamma }
{\beta }} \right)P_5  + \frac{{\dot{\overline{ P}}_B }}
{{12M^{3}}}\,,\label{chi}
\end{equation}
where the defining equation for $\chi$ is
\begin{equation}
\frac{\ddot{a}}{a}=-\chi-(3w+2)\beta\rho^2-(3w+1)\gamma\rho\,.
\end{equation}
The first of these equations (\ref{feq}) is analogous to the Friedmann equation of standard cosmology.
The dark energy field accounts for contributions to the Friedmann equation which do not come from ordinary brane matter. In the absence of bulk content, it is equal to a dark radiation term, a consequence of the presence of the extra dimension. When $\bar P_{B}$ or $P_5$ are non-zero, it also receives contributions from these terms. Thus the origin of $\chi$ can be both geometric and matter-related. 

We are interested in scenarios where the energy density of the brane is much lower than its tension, that is $\rho << \sigma$. We can then disregard terms quadratic in $\rho$ and, thus, we get the simplified cosmological equations
\begin{equation}
\left( {\frac{{\dot a}}
{a}} \right)^2 \, =\,2\gamma \rho  - \frac{k}
{{a^2 }} + \chi  - \frac{\overline{P}_B }
{{12M^{3}}}\,,
\label{coeq1}
\end{equation}
\begin{equation}
\dot \chi  + 4\frac{{\dot a}}
{a}\left( \chi  - \frac{\overline{P}_B }
{24M^{3}} \right)\,=4\gamma P_5   + \frac{{\dot {\overline{P}}_B }}
{{12M^{3}}}\,.
\end{equation}
and, of course, the continuity equation
\begin{equation}
 \dot \rho  + 3\rho \frac{{\dot a }}
{{a}}\left( {1 + w} \right) =  - 2P_5\,.
\label{coeq3}
\end{equation}

The functions $\overline{P}_{B}$, $P_5$ are functions of time 
corresponding to the values of $\overline{P}_B(y,t)$ and $P_5(y,t)$ on the brane. The energy-momentum conservation
$\nabla_MT^M_{\,\,N}=0$ cannot fully determine $\overline{P}_{B}$ and $P_5$ and a particular model of the bulk matter 
is required \cite{BT}. Here we are going to consider a general ansatz for the bulk pressures (see also \cite{CGW})
\begin{equation}
\overline{P}_B\,=\,D\,a^{\nu}\,\,,\,\,\,\,P_5\,=\,F\,\left(\frac{\dot{a}}{a}\right)\,a^{\mu}\,.{\label{Ansatz}}
\end{equation}
where $F$ and $D$ are constant parameters. For this choice of pressures, the equation for $\chi$ can be solved exactly to give
\begin{equation}
\chi\,=\,\frac{{\cal{C}}}{a^4}\,+\,2\delta\frac{( \nu+2)}{(\nu+4)}a^{\nu}\,+\,\frac{4F\gamma}{(\mu+4)}a^{\mu}\,,{\label{eqxi}}
\end{equation}
where we have defined $\delta\equiv D/24M^3$. Similarly, we can proceed with the integration of (\ref{coeq3}) and get
\begin{equation}
\rho\,=\,\frac{\tilde{\cal{C}}}{a^{3(1+w)}}\,-\frac{2F}{\left[3(1+w)+\mu\right]}a^{\mu}\,.{\label{eqrho}}
\end{equation}
Substituting (\ref{eqxi}) and (\ref{eqrho}) into the Friedmann equation (\ref{coeq1}), we can write it in a conventional form as 
\begin{equation}
\left(\frac{\dot{a}}{a}\right)^2\,+\,\frac{k}{a^2}\,=\,\frac{8\pi}{3}G_N\,\rho_{eff}\,,{\label{Fried}}
\end{equation}
where $G_N=3\gamma/4\pi=3\sigma/4\pi(24M^3)^2$ is the $4D$ Newton's constant and the {\textit{effective energy density}} $\rho_{eff}$ stands for
\begin{equation}
\rho_{eff}\,=\,\frac{\tilde{\cal{C}}}{a^{3(1+w)}}\,+\,\frac{{\cal{C}}/2\gamma}{a^4}\,-\frac{2\delta}{\gamma(\nu+4)}a^{\nu}\,
+\,\frac{2(3w-1)F}{(\mu+4)\left[3(1+w)+\mu\right]}a^{\mu}\,.{\label{rhoeff}}
\end{equation}
At late times, the term with the highest power in $a$ will dominate and we will get a scale factor $a(t) \sim t^{-\frac{2}{max(\mu,\nu)}}$. Acceleration occurs whenever $max(\mu,\nu)>-2$.

Using the already mentioned effective energy density, we can derive an expression for the\textit{ effective equation of state parameter} using the prescription\cite{LJ}
\begin{equation}
w_{eff}^{(D)}\,=\,-1\,-\frac{1}{3}\frac{d\ln (\delta H^2)}{d\ln a}\,,{\label{eosp}}
\end{equation}
where $\delta H^2=H^2/H_0^2-\Omega_ma^{-3}$ accounts for all terms in the Friedmann equation not related to the brane matter, for which we have taken $w=0$. Using as our variable the redshift parameter $z=\frac{a_0}{a(t)}-1$, $w_{eff}$ has the general form
\begin{equation}
w_{eff}  =  - 1 +
 \frac{1}{3}\left(\frac{{\nu A + \mu B\left( {z + 1} \right)^{\nu  - \mu } }}{{ - A - B\left( {z + 1} \right)^{\nu  - \mu }
 }}\right)\,,
\label{weff}
\end{equation}
where the parameters $A$ and $B$ depend on the bulk pressures 
\begin{equation}
A \equiv \frac{{2\delta }}{{\gamma \left( {\nu  + 4} \right)}} = \frac{{48DM^3 }}{{\left( {\nu  + 4} \right)\sigma }},\,\,\,
B  \equiv \frac{{2F}}{{\left( {\mu  + 3} \right)\left( {\mu  + 4} \right)}}\,.
\end{equation}
We have assumed that the dark radiation parameter ${\cal{C}}$ is small and that, at late times, the dark
 radiation term $\sim \frac{\mathcal{C}/2 \gamma}{a^4}$ is negligible with respect to the contribution of the 
pressures $\bar{P}_B$ and $P_5$. Thus, we have ignored this term in our expression for $w_{eff}$.

Equation (\ref{weff}) allows us to examine under which conditions we can have an effective 
equation of state which can exhibit $w_{eff}=-1$ crossing, as is suggested by recent observational 
data. In particular, we want to investigate what are the requirements that our model's parameters have
to comply with, in order to get a qualitatively correct behavior for the
evolution of $w_{eff}$ at 
late times. Current observational findings indicate that at the moment ($z=0$)
we are at a stage of
accelerated expansion with  $\left| q \right|$ of order unity and effective
equation of state $w_{eff}=-1.21$. 
The $w_{eff}=-1$ crossing seems to have occurred at redshifts $z=0.2$. Therefore
$w_{eff}$ seems to increase with $z$, 
starting with a value of about $-1.21$ at $z=0$ and crossing the $w_{eff}=-1$ line
about $z=0.2$. Our goal is to obtain
constraints on the values of our parameters $A$, $B$, $\mu$ and $\nu$ and see
for which combinations we can have
a $w_{eff}$ with the desired temporal profile, while having accelerated
expansion. The corresponding late time deceleration parameter (where we have
neglected the contribution from brane matter as well as the dark radiation term)
is
\begin{equation}
q \equiv  - \frac{1}{{H^2 }}\frac{{\ddot a}}{a} =  - \frac{{\left( {\nu  + 2} \right)A + \left( {\mu  + 2} \right)B\left( {z + 1} \right)^{\nu  - \mu } }}{{2A + 2B\left( {z + 1} \right)^{\nu  - \mu } }}\,.
\end{equation}

By inspecting equation (\ref{weff}), we first notice that the denominator is 
the effective energy density at late times and thus it has to be positive at the
interval of $z$ we are interested in (in this discussion we focus on $0\le z\le
1$). At $z=0$ this condition reduces to $A+B<0$. It is, thus, obvious that $A$
and $B$ cannot be both positive. 
Taking the derivative of (\ref{weff}), we obtain
\begin{equation}
\frac{{dw_{eff} }}{{dz}} = AB\frac{{\left( {\nu  - \mu } \right)^2 \left( {z + 1} \right)^{\nu  - \mu  - 1} }}{{\left( {A + B\left( {z + 1} \right)^{\nu  - \mu } } \right)^2 }}\,.
\label{der}
\end{equation}
All terms in the right hand side of (\ref{der}) are positive 
in the interval $0\le z\le 1$, except from the product $AB$. In order to have 
an increasing $w_{eff}$, $A$ and $B$ must have the same sign, i.e. they must both be negative ($A,B<0$).
Next, we want $w_{eff}$ to assume a value less than $-1$ at $z=0$. 
In order to achieve this we must have $\nu A + \mu B < 0$. 
There are four possible combinations:

1) $\mu,\nu>0$: The constraint is satisfied.

2) $\mu>0$, $\nu<0$: $\nu A + \mu B < 0$ when $\frac{\nu }{\mu } >  - \frac{B}{A}$.

3) $\mu<0$, $\nu>0$: $\nu A + \mu B < 0$ when $\frac{\nu }{\mu } <  - \frac{B}{A}$.

4) $\mu,\nu<0$: The constraint cannot be satisfied for $A$ and $B$ negative.

So, the first three cases seem to be allowed. However, 
a further constraint comes from the fact that we want $w_{eff}=-1$ for $z=0.2$. 
This reduces to the relation
\begin{equation}
\frac{\nu }{\mu } =  - \frac{B}{A}(1.2)^{\nu  - \mu } \,.
\label{eq1}
\end{equation}
In the case (1), with both $\mu$ and $\nu$ positive, the left hand side of
(\ref{eq1}) is positive, while the right hand side is negative and thus it is
ruled out. Cases (2) and (3) can satisfy this equation without providing any
additional constraints for the powers $\mu$ and $\nu$. Thus we conclude that the
only choices of parameters which are in accordance with the general profile for
the evolution of $w_{eff}$ for $0 \le z \le 1$ are:

1) $A,B<0$ with $\mu>0$, $\nu<0$ and $\frac{\nu }{\mu } >  - \frac{B}{A}$.

2) $A,B<0$ with $\mu<0$, $\nu>0$ and $\frac{\nu }{\mu } <  - \frac{B}{A}$.

Since one of the powers has to be positive, the condition $max(\mu,\nu)>-2$ is
always satisfied and we also get the desired accelerated expansion for late
times. We also see that the ratio of bulk pressure parameters $A$ and $B$
determines the ratio of the two powers. A large difference between  the initial
pressures $D$ and $F$ would result in an accordingly large disparity between
$\mu$ and $\nu$. 
{This is demonstrated in Figure 1, where we give a graph of the
deceleration parameter $(q)$ and $w_{eff}$, for a set of values that satisfy the
above constraints. For this choice, we see  that there is an accelerated
expansion at late times and $w_{eff}$ has a realistic temporal profile.

\begin{figure}[t]
 \centering
     \begin{minipage}[c]{0.8\textwidth}
    \centering \includegraphics[width=\textwidth]{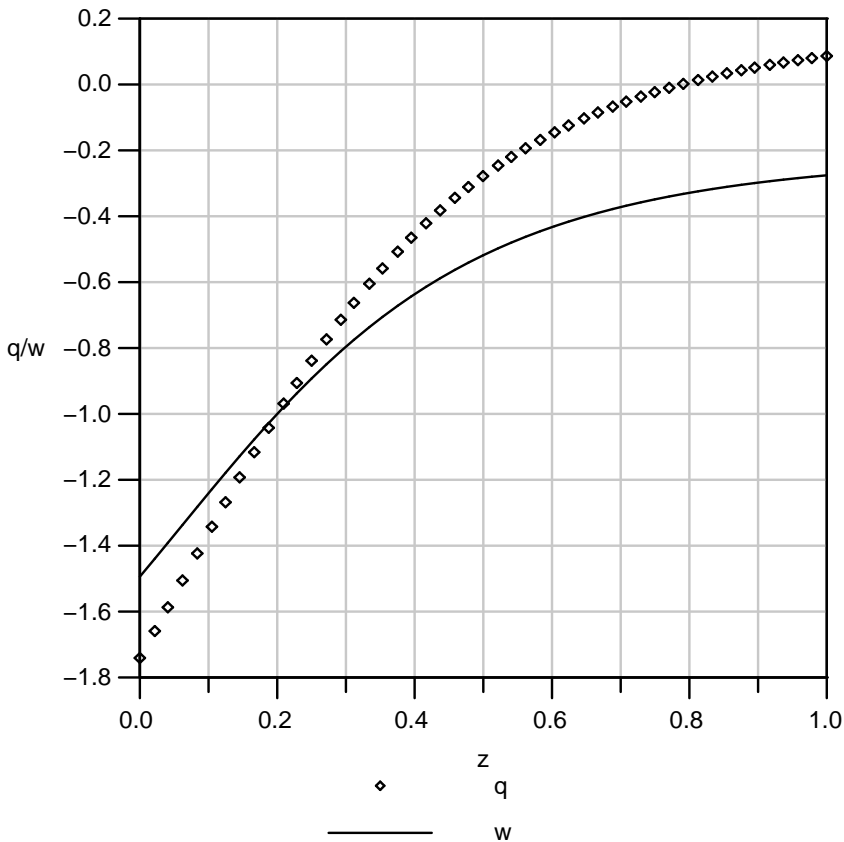}
     \end{minipage}
    \label{figure1}
    \caption{Graph of the deceleration parameter q and $w_{eff}$ for $A=-1$,
$B=-2$, $\mu=2$, $\nu=-2$.}
 \end{figure}

The requirement that $A$ and $B$ must be negative has a direct impact on the
sign of parameters $D$ and $F$ and the corresponding behavior of the underlying
bulk content. In order to have $B<0$ with $\mu<-4$ or $\mu>-3$, we must
have $P_5<0$, which means we have energy inflow from the bulk. When
$-4<\mu<-3$, we get $P_5>0$ and energy
outflow from the brane. Similarly, for $A<0$, we get $\bar P_B<0$ if $\nu>-4$.
This means that the bulk content 
must have negative pressure. This resembles some sort of five-dimensional
quintessence field (like a scalar field). For $\nu<-4$, the pressure $\bar
P_B$ becomes positive and
we can assume the bulk matter to behave as a regular relativistic fluid.
Apparently, we cannot have both pressures positive, since then we would have two
negative powers, which is excluded from the constraints obtained above. Notice
also that we have managed to achieve $w_{eff}=-1$
crossing without the use of any bulk matter which violates energy
conditions, like phantom fields.
The presence of the dark radiation term was also not necessary to get the
desired phenomenon, which manifests itself as a result of the interplay between the bulk pressure and
the brane-bulk 
energy exchange terms. We, thus, have a description of dark energy in 
terms of a bulk energy-momentum tensor, which can drive accelerated expansion
in the present era and provide a qualitatively correct picture for the
evolution of the equation of state at late times.

As a further application of our model, we can show that with proper choice of parameters 
for the bulk pressure, we can simulate a Chaplygin gas, which is another candidate for quintessence.
A Chaplygin gas is a perfect four-dimensional fluid which obeys the equation of
state 
\begin{equation}
p =  - \frac{A}{\rho }\,.
\end{equation}
Assuming energy conservation in the form
\begin{equation}
d\left( {\rho a^3 } \right) =  - pd\left( {a^3 } \right)\,,
\end{equation}
it is easy to show that for small $a$, i.e. early times,  $\rho \propto a^{-3}$. This is the 
behavior we expect from dust matter with $w=0$. For large $a$ or late times, $\rho  =  - p$, which 
is the equation of state for cosmological constant. The Chaplygin gas behaves as dark matter at early times and as dark energy (cosmological constant) at late times and is a simple example of unified treatment of these two constituents. It is easy to see that the $w_{eff}$ in (\ref{weff}) can replicate this behavior and thus the bulk content can play a role similar to that of a Chaplygin gas. We first want the effective equation of state parameter to be equal to $-1$ at late times ($z=0$). This can be satisfied for $\nu A + \mu B = 0$, or
\begin{equation}
\frac{\nu }{\mu } =  - \frac{{\rm B}}{A}\,.
\end{equation}
Additionally, we want $w_{eff}=0$ for $z>>1$. Depending on which is the highest power, we must have either $\mu=-3$ for $\nu-\mu>0$, or $\nu=-3$ for $\nu-\mu<0$. The corresponding conditions are then:

1) $\mu=-3$, $\nu  = 3\frac{B}{A}$.

2) $\nu=-3$, $\mu  = 3\frac{A}{B}$.

For the above choice of parameters, we can reproduce the effects of a Chaplygin gas using only bulk matter content.

Summarizing, we dealt with the problem of simulating the
properties of dark energy in the context of brane world cosmology. We showed
that by assuming the existence of an extra dimension and bulk matter, 
whose energy-momentum tensor is of the form (\ref{Ansatz}), we can reproduce
the observed accelerated expansion of the universe and the equation of state for
the dark energy. We analyzed in detail the constraints our model 
parameters have to obey in order to get a realistic profile for the evolution
of $w_{eff}$ for small $z$. The crossing of the $w_{eff}=-1$ line occurs naturally,
without invoking exotic forms of bulk or brane matter that violate
energy conditions. Our results also have the advantage of not depending on the
dark radiation term and they consequently do not fix the sign of the
$\mathcal{C}$ constant. We also showed that this model can reproduce predictions
similar to those of other dark energy candidates, in particular the Chaplygin
gas.We cannot, at the moment, fully
justify
the ansatz (\ref{Ansatz}) in terms of a phenomenological bulk fluid. As it was
shown in \cite{BT}, this is due to the unknown spatial derivatives of the bulk
pressures that appear in the equation of energy-momentum conservation for the
bulk fluid. A specific model for the bulk matter is needed, in order to
eliminate this extra freedom. Such a realistic model for the bulk matter, which
may yield the desired behavior, is currently under investigation.

{\textbf{ Acknowledgments.}}  
A. D. and C. B. wish to thank S. Nesseris for useful discussions. This research was co-funded by the European Union
in the framework of the Program $\Pi Y\Theta A\Gamma O PA\Sigma-II$ 
of the {\textit{``Operational Program for Education and Initial Vocational Training"}} ($E\Pi EAEK$) of the 3rd Community Support Framework
of the Hellenic Ministry of Education, funded by $25\% $ from 
national sources and by $75\%$ from the European Social Fund (ESF). C. B.
acknowledges also an {\textit{Onassis Foundation}} fellowship.


\begin{thebibliography}{99}

\bibitem{R} A. G. Riess et al., Astron. J. {\textbf{116}}, 1009 (1998); S. Perlmutter et al., Astrophys. J. {\textbf{517}}, 565 (1999);
 A. G. Riess et al.Astrophys.J. {\textbf{607}}, 665 (2004).
\bibitem{WMAP} D. N. Spergel, et al., WMAP Three Year Results: Implications for Cosmology, astro-ph/0603499.

\bibitem{PHANT} R. R. Calwell, Phys. Lett. B {\textbf{545}}, 23 (2002); R. R. Caldwell, M. Kamionkowski and N. N. Weinberg, Phys. Rev. Lett. {\textbf{91}}, 071301
(2003); J. M. Cline, S. Y. Jeon and G. D. Moore, Phys. Rev. D {\textbf{70}}, 043543 (2004).
 
\bibitem{QUINT} R. R. Caldwell, R. Dave and P. J. Steinhardt, Phys. Rev. Lett. {\textbf{80}}, 1582 (1998); P. J. E. Peebles and A. Vilenkin, Phys. Rev. D
{\textbf{59}}, 063505 (1999); P. J. Steinhardt, L. M. Wang and I. Zlatev, Phys. Rev. D {\textbf{59}}, 123504 (1999); M. Doran and J. Jaeckel, Phys. Rev. D
{\textbf{66}}, 043519 (2002); A. R. Liddle, P. Parson and J. D. Barrow, Phys. Rev. D {\textbf{50}}, 7222 (1994).

\bibitem{WEI}
  H.~Wei, R.~G.~Cai and D.~F.~Zeng,
  Class.\ Quant.\ Grav.\  {\bf 22}, 3189 (2005)
  [arXiv:hep-th/0501160].

\bibitem{KMP} 
  A.~Y.~Kamenshchik, U.~Moschella and V.~Pasquier,
  Phys.\ Lett.\ B {\bf 511} (2001) 265
  [arXiv:gr-qc/0103004].

\bibitem{SS}
  V.~Sahni and A.~Starobinsky,
  arXiv:astro-ph/0610026.

\bibitem{AHDD} N. Arkani-Hamed, S. Dimopoulos and G. R. Dvali, Phys. Lett. B {\textbf{ 429}} (1998) 263; I. Antoniadis, N. Arkani-Hamed, S. Dimopoulos and G. R.
Dvali, Phys. Lett. B {\textbf{436}} 257 (1998).

\bibitem{DGP} G. R. Dvali, G. Gabadadze and M. Porrati, Phys. Lett. B {\textbf{485}}, 208 (2000); G. R. Dvali, G. Gabadadze, M. Kolanovic and F. Nitti, Phys. Rev. D
{\textbf{64}} 084004 (2001).

\bibitem{RS} L. Randall and R. Sundrum, Phys. Rev. Lett. {\textbf{83}} (1999) 3370; Phys. Rev. Lett. {\textbf{83}} (1999) 4690.

\bibitem{BDL} P. Binetruy, C. Deffayet and D. Langlois, Nucl. Phys. B
{\textbf{565}} 269 (2000); P. Binetruy, C. Deffayet, 
U. Ellwanger and D. Langlois, Phys. Lett. B {\textbf{447}} 285 (2000).

\bibitem{KKTTZ} E. Kiritsis, G. Kofinas, N. Tetradis, T. N. Tomaras and V. Zarikas, JHEP {\textbf{0302}} (2003) 035; E. Kiritsis, N. Tetradis and T. N. Tomaras, 
JHEP {\textbf{0203}} (2002) 019; P. S. Apostolopoulos and N. Tetradis, Phys. Rev. D {\textbf{71}} 043506 (2005); P. S. Apostolopoulos and N. Tetradis, Phys. Lett. B
{\textbf{633}} 409 (2006); E. Kiritsis, JCAP {\textbf{0510}} 014 (2005); K. I. Umezu, K. Ichiki, T. Kajino, G. J. Mathews, R. Nakamura and M. Yahiro, Phys. Rev. D
{\textbf{73}} 063527 (2006).

\bibitem{BT} 
  C.~Bogdanos and K.~Tamvakis,
  arXiv:hep-th/0609100.


\bibitem{COSMO} T. Padmanabhan, Phys. Rept. {\textbf{380}}, 235 (2003); V. Sahni and A. A. Starobinsky, Int. J. Mod. Phys. D {\textbf{9}}, 373 (2000); S. M. Carroll,
Living Rev. Rel. {\textbf{4}}, 1 (2001); S. Weinberg, Rev. Mod. Phys. {\textbf{61}}, 1 (1989).

\bibitem{A} I. Antoniadis, Phys. Lett. B {\textbf{246}}, 377 (1990).


\bibitem{SMS} T. Shiromizu, K. Maeda and M. Sasaki, Phys. Rev. D {\textbf{62}} (2000) 024012.

\bibitem{Pal:2006hg}
  S.~Pal,
  Phys.\ Rev.\ D {\bf 74} (2006) 024005
  [arXiv:gr-qc/0606085].

\bibitem{NP1}
  S.~Nesseris and L.~Perivolaropoulos,
  Phys.\ Rev.\ D {\bf 70}, 043531 (2004)
  [arXiv:astro-ph/0401556];   S.~Nesseris and L.~Perivolaropoulos,
  [arXiv:astro-ph/0610092]. 
   
\bibitem{CGW} R.G. Cai, Y. Gong and B. Wang, JCAP 0603 (2006); P. S. Apostolopoulos and N. Tetradis, hep-th/0604014.



\bibitem{LJ} E. V. Linder and A. Jenkins, Mon. Not. Roy. Astron. Soc. {\textbf{346}}, 573 (2003).


\end{thebibliography}
\end{document}